\documentclass[10pt]{iopart}
\usepackage[utf8]{inputenc}
\usepackage{graphicx}
\usepackage{booktabs}
\usepackage{multirow}
\usepackage{cite}
\usepackage{amssymb}
\usepackage[hidelinks]{hyperref}
\begin{document}
\frenchspacing

\title{Decoding ECoG signal into 3D hand translation using deep learning}

\author{Maciej Śliwowski$^{1, 2}$, Matthieu Martin$^1$, Antoine Souloumiac$^2$, Pierre Blanchart$^2$ and Tetiana Aksenova$^1$}

\address{1 Univ. Grenoble Alpes, CEA, LETI, Clinatec, F-38000 Grenoble, France}
\address{2 Université Paris-Saclay, CEA, List, F-91120, Palaiseau, France}
\ead{\mailto{maciej.sliwowski@cea.fr},\mailto{tetiana.aksenova@cea.fr}}
\vspace{10pt}
\begin{indented}
\item[]September 2021
\end{indented}

\begin{abstract}
\textit{Objective.}
Motor brain-computer interfaces (BCIs) are a promising technology that may enable motor-impaired people to interact with their environment. BCIs would potentially compensate for arm and hand function loss, which is the top priority for individuals with tetraplegia.
Designing real-time and accurate BCI is crucial to make such devices useful, safe, and easy to use by patients in a real-life environment.
Electrocorticography (ECoG)-based BCIs emerge as a good compromise between invasiveness of the recording device and good spatial and temporal resolution of the recorded signal. However, most ECoG signal decoders used to predict continuous hand movements are linear models. These models have a limited representational capacity and may fail to capture the relationship between ECoG signal features and continuous hand movements. Deep learning (DL) models, which are state-of-the-art in many problems, could be a solution to better capture this relationship.

\textit{Approach.}
In this study, we tested several DL-based architectures to predict imagined 3D continuous hand translation using time-frequency features extracted from ECoG signals. The dataset used in the analysis is a part of a long-term clinical trial (ClinicalTrials.gov identifier: NCT02550522) and was acquired during a closed-loop experiment with a tetraplegic subject. The proposed architectures include multilayer perceptron (MLP), convolutional neural networks (CNN), and long short-term memory networks (LSTM). The accuracy of the DL-based and multilinear models was compared offline using cosine similarity.

\textit{Main results.}
Our results show that CNN-based architectures outperform the current state-of-the-art multilinear model. The best architecture exploited the spatial correlation between neighboring electrodes with CNN and benefited from the sequential character of the desired hand trajectory by using LSTMs. Overall, DL increased the average cosine similarity, compared to the multilinear model, by up to 60\%, from 0.189 to 0.302 and from 0.157 to 0.249 for the left and right hand, respectively.

\textit{Significance.}
This study shows that DL-based models could increase the accuracy of BCI systems in the case of 3D hand translation prediction in a tetraplegic subject.

\end{abstract}
\vspace{2pc}

\ioptwocol

\section{Introduction}
Brain-computer interfaces (BCIs) enable people to interact with their environment using a direct connection between their brain and external effectors. BCI provides humans with an alternative way to interact with their surroundings and could substitute lost motor function, for example, in tetraplegic persons. Several communication paradigms have been designed that are suitable for a broad range of tasks, like keyboard typing or binary decision making. In this study, we focus on motor imagery (MI) based BCI, which can be used for continuous and asynchronous \cite{guger_motor_imagery} control of, for example, exoskeleton \cite{benabid_exoskeleton_2019}. Such devices could improve a paralyzed person's quality of life by offering them a certain autonomy, especially for the tasks of everyday life. In the MI BCI paradigm, subjects imagine movements that cause changes in brain activity. Traces of this activity can be recorded, usually in the form of electrical potential or magnetic field, and decoded into external device commands.
We focused on upper-limb movements imagination because regaining arm and hand motor functions are at the top of the priority list in individuals with tetraplegia \cite{simpson_hand_needed, anderson_hand_needed}.

A typical BCI system consists of a brain signal recording device, a feature extraction step, a decoder that translates features into actions understandable by a computer, and finally, a device that executes the commands from the decoder \cite{shih_brain-computer_2012}.

The first part of the BCI system is a brain signal recording device. Acquired signal characteristics strongly depend on the recording device. Recording methods can be separated into invasive and noninvasive. The most invasive devices are intracortical microelectrode arrays (MEA) that are placed inside the brain tissue \cite{hochberg_reach_2012, collinger_high-performance_2013, wodlinger_ten-dimensional_2015}. Electrocorticography (ECoG) is a less invasive method that records signal from the brain surface \cite{benabid_exoskeleton_2019, nakanishi_decoding_2014, flint_continuous_2017}. The most common noninvasive recording method used for BCI is electroencephalography (EEG), monitoring the electrical activity of the brain using a set of electrodes placed over the scalp \cite{lawhern_eegnet_2018, schwarz_decoding_2018, mondini_continuous_2020, korik_decoding_2018}.
This study focuses on ECoG-based BCI because it has better signal characteristics than noninvasive methods while decreasing the risk connected to implantation and biocompatibility issues compared to MEA. ECoG signal is also more stable in time \cite{larzabal_long-term_2021}. Therefore, ECoG-driven BCI emerges as a promising tool for neuroprosthesis development.

The next step after signal recording is to extract features from the signal. Those features represent brain activity in a form that a decoder can exploit. This transformation depends on the experimental paradigm and the type of recorded signals.
The most common and effective representation of ECoG and EEG signals for MI BCI are time-frequency features. They contain information about power time course in several frequency bands \cite{schalk_decoding_2007, chin_identification_2007, sanchez_extraction_2008, liang_decoding_2012, du_decoding_2018, benabid_exoskeleton_2019}, or focus only on low-frequency component (LFC)/Local Motor Potential (LMP) \cite{schalk_decoding_2007, pistohl_prediction_2008, acharya_electrocorticographic_2010}.

The next step in a typical BCI system is a decoder, translating brain signal features into BCI commands. Decoding performance is crucial for the quality of interaction between humans and computers. Higher accuracy of decoding improves correctness and speed of interaction. Most current decoders use machine learning (ML) algorithms to predict BCI commands. ML models are data-driven solutions that utilize data to estimate the decoder's parameters to predict the target variable accurately. In most cases, supervised learning approaches are used---models are trained using data collected during a calibration phase with pre-defined targets to reach.

Predicting BCI commands based on brain signals is a challenging task due to several limitations originating from the nature of the application. Recorded brain signals have a strong component of noise which is generated by spontaneous brain activity not related to the task. The recorded signal is nonstationary in time (intra-subject variability) which often makes models valid only for a limited time and requires decoder retraining. BCI experiments are monotonous and time-consuming, so ML models have to be trained with small datasets to reduce the time needed for calibration. Another important constraint is the need for real-time decoding---the whole system should produce several movement predictions per second.
In the case of tetraplegic subjects, the real intention of the patient is not known to the experimenters. Even if a subject is provided with explicit instruction, actual imagination patterns can be affected by several factors, e.g., attention and fatigue levels or trajectory corrections.

Several types of algorithms were used to decode brain signals. Majority of studies use 'conventional' ML techniques \cite{chauhan_review_2018} for decoding of hand movement. In the case of intracortical recordings, linear models, e.g. Ridge linear regression \cite{collinger_high-performance_2013, wodlinger_ten-dimensional_2015} and Kalman filtering \cite{hochberg_reach_2012, ajiboye_restoration_2017} were applied to decode continuous imagined hand trajectories. For ECoG signal decoding, many studies focus on discrete decoding of hand gestures/movements or fingers flexion, typically using standard classifiers. This includes linear discriminant analysis (LDA) classifier \cite{pistohl_decoding_2012, milekovic_online_2012, gunduz_differential_2016, kapeller_single_2014} or support vector machine (SVM) \cite{li_gesture_2017, yanagisawa_electrocorticographic_2012, wang_human_2009, liu_decoding_2010, spuler_decoding_2014, xu_fangzhou_decoding_2020}, as well as other methods like naive Bayes classifier \cite{chestek_hand_2013} or spatiotemporal template matching \cite{bleichner_give_2016, branco_decoding_2017}. Some of them were combined with additional data dimensionality reduction methods like principal component analysis (PCA) \cite{wang_human_2009, liu_decoding_2010} or common spatial pattern (CSP) \cite{kapeller_single_2014}. Another group of studies demonstrated prediction of continuous targets---2D and 3D hand movements and fingers flexion---mostly using linear models including linear regression and its variants \cite{schalk_decoding_2007, nakanishi_prediction_2013, nakanishi_mapping_2017, kubanek_decoding_2009, flamary_decoding_2012, liang_decoding_2012, weixuan_chen_logistic_weighted_2014} partial least squares (PLS) \cite{bundy_decoding_2016}, recursive exponentially weighted n-way partial least squares (REW-NPLS) \cite{benabid_exoskeleton_2019}, and Kalman filtering \cite{pistohl_prediction_2008, silversmith_plug-and-play_2020, kellis_decoding_2012}.

Another group of machine learning algorithms are deep learning (DL) based methods. The core idea behind DL is stacking series of nonlinear transformations---layers---that create a deep structure extracting complex representations. Each layer consists of multiple trainable units (traditionally called neurons) performing basic operations. Although DL-based methods got more attention and demonstrated their usefulness in a variety of problems, e.g., in computer vision \cite{krizhevsky_imagenet_2012} or natural language processing \cite{devlin-etal-2019-bert}, the vast majority of ECoG based studies use 'conventional' machine learning algorithms, mainly linear models. However, some attempts were made to use DL for intracortical, ECoG, and EEG signals analysis in humans. Almost all reported solutions use deep learning only for classification tasks. Studies analyzing intracortical recordings used recurrent neural networks (RNN) based autoencoders \cite{pandarinath_inferring_2018} as well as CNN \cite{golshan_lfp-net_2020}, and a combination of RNN and CNN \cite{schwemmer_meeting_2018} to recognize several MI classes using neuronal firing rates or time-frequency representation of single units activation.

In the EEG domain, many works evaluated DL-based solutions for signal classification including both extracted features and raw signals approaches. Most common architectures employed CNN using hand-crafted features \cite{roy_deep_2020, tabar_novel_2017, tang_single-trial_2017, xu_wavelet_2019} or raw signals \cite{schirrmeister_deep_2017, lawhern_eegnet_2018, dose_end_end_2018, huang_deep_2019, dai_hs-cnn_2020} as inputs, long short-term memory networks (LSTM) \cite{ma_improving_2018, wang_lstm-based_2018} and mix of CNN and LSTM analyzing time-frequency features \cite{bashivan_learning_2016, zhang_novel_2019, garcia-moreno_cnn-lstm_2020}. There are only a few studies that used DL approaches to decode ECoG signals. In particular, LSTM were used to classify fingers activation \cite{du_decoding_2018, elango_sequence_2017} and 3 different hand gestures \cite{pan_rapid_2018} from time-frequency features. Rashid \etal \cite{rashid_electrocorticography_2020} discriminated tongue and hand movements from raw signal using LSTM.

To our best knowledge, only two studies considered DL-based approaches to predict continuous variables from brain signals. Park and Kim \cite{park_estimation_2019} used a 1-layer LSTM to predict 2D hand movements (direction and speed) in the case of monkeys performing center-out-task. Input features consisted of neuronal firing rates extracted from intracortical recordings. For ECoG signals, Xie \etal \cite{xie_decoding_2018} predicted continuous flexion and extension of 5 fingers using end-to-end DL with four spatial/temporal convolutional layers as feature extractors and one LSTM layer to predict fingers activation. Each 1D finger activation was predicted using a separated neural network. Both works \cite{park_estimation_2019, xie_decoding_2018} predicted actual hand/fingers movements trajectories. These allow creating an explicit mapping between brain activity and trajectory, removing the uncertainty introduced in the case of only imagined movements. In both cases, recordings were performed with non-disabled patients executing overt movements, so the signals were not affected by feedback control and the patient's movement corrections.

Continuous decoding is the only way to provide paralyzed patients with normal-like control of complex neuroprosthetics. The maximum level of continuous control achieved using DL-based approaches, until now, is 2D hand trajectory using intracortical recordings. ECoG continuous decoding with DL was only done to predict 1D finger flexion. Otherwise, multilinear models have been used to predict up to 3D continuous movements \cite{benabid_exoskeleton_2019, bundy_decoding_2016, nakanishi_prediction_2013, nakanishi_mapping_2017}.

As DL achieved state-of-the-art performance in solving several complex problems from different domains, in this work, we propose and evaluate several DL architectures to predict 3D continuous hand movements. For the first time, we show that DL can efficiently predict complex, high-dimensional, upper-limb translation trajectories from ECoG signals. The dataset used for the comparison was recorded in a closed-loop experiment in which adaptive multilinear REW-NPLS \cite{eliseyev_recursive_2017} models were used. Recordings were performed with a tetraplegic patient within more than 200 days. DL-based models obtained better performance in an offline comparison than multilinear models optimized by REW-NPLS.

\section{Methods}

\subsection{Clinical trial and patient} \label{sec:dataset}
This study was a part of a clinical trial "BCI and Tetraplegia" (ClinicalTrials.gov identifier: NCT02550522) approved by French authorities: Agency for the Safety of Medicines and Health Products (Agence nationale de sécurité du médicament et des produits de santé---ANSM) with the registration number: 2015-A00650-49 and the ethical Committee for the Protection of Individuals (Comité de Protection des Personnes---CPP) with the registration number: 15-CHUG-19. Clinical trial details are described by Benabid \etal \cite{benabid_exoskeleton_2019}.

The subject involved in this study was a 28-year-old right-handed man with tetraplegia caused by spinal cord injury \cite{benabid_exoskeleton_2019}.

\subsection{Experiment and dataset}

\begin{figure}
	\centering \includegraphics[width=\columnwidth]{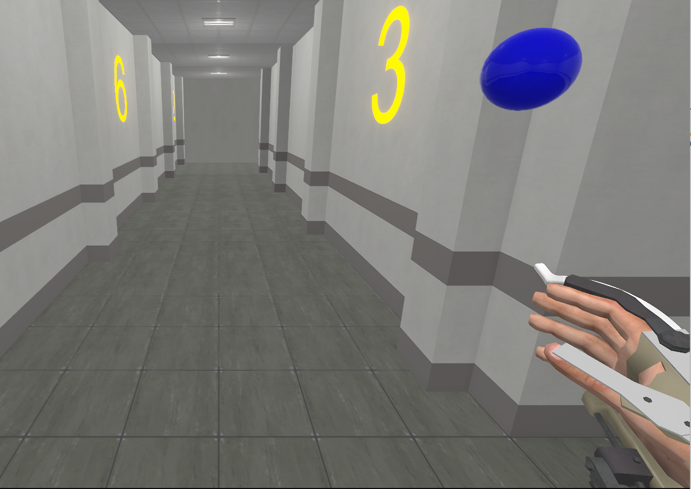}
	\caption{Screenshot from the virtual environment displaying the hand of the avatar and the target.} \label{fig:experiment-ve}
\end{figure}

During the experiment, the patient had control over an avatar with a first-person view in a virtual environment. He could control one state out of 5: idle, left or right hand 3D translation, and left or right wrist rotation. This article focuses only on the left and right hand translation. In that case, spherical targets (10 cm diameter) were displayed in space and the patient task was to reach them successively (Figure \ref{fig:experiment-ve}). To control the virtual avatar, the patient used an MI strategy developed in the previous experiments of the clinical trial. This strategy consisted of imagined arm, wrist, and finger movements. The patient was instructed to maintain a constant imagination strategy through the experiments.
 	 
During the recordings, the hand movement predictions were performed by multilinear models obtained by REW-NPLS \cite{eliseyev_recursive_2017}. Separate models for each hand were incrementally calibrated during the six first sessions and used without re-calibration for the following 37 sessions. The data acquired during these sessions was used to perform simulations of models offline training to evaluate their potential benefits for this application. To ensure equal conditions for further comparisons, we decided to retrain all models on the same datasets.

Targets were placed in 28 (LH) and 23 (RH) positions during the experiments (see \ref{app:targets-positions} for targets positions visualization). The number of trials and minutes of the recorded signal are given for both hands and the calibration and test datasets in Table \ref{tab:data}.

\begin{table}[]
	\caption{Datasets size in the number of trials and length of the recordings.}
	\label{tab:data}
	\centering
	\resizebox{\columnwidth}{!}{%
		\begin{tabular}{ccccc}
			\br
			& \multicolumn{2}{c}{Left hand}                                            & \multicolumn{2}{c}{Right hand}                                           \\ \mr
			& Trials & \begin{tabular}[c]{@{}c@{}}Duration [min]\end{tabular} & Trials & \begin{tabular}[c]{@{}c@{}}Duration [min]\end{tabular} \\ \mr
			Calibration & 174    & 42                                                              & 164    & 39                                                              \\
			Test        & 691    & 314                                                             & 649    & 327                                                             \\ \br
		\end{tabular}%
	}
\end{table}

\subsection{Signal recording}
\begin{figure}
	\centering \includegraphics[width=0.6\columnwidth]{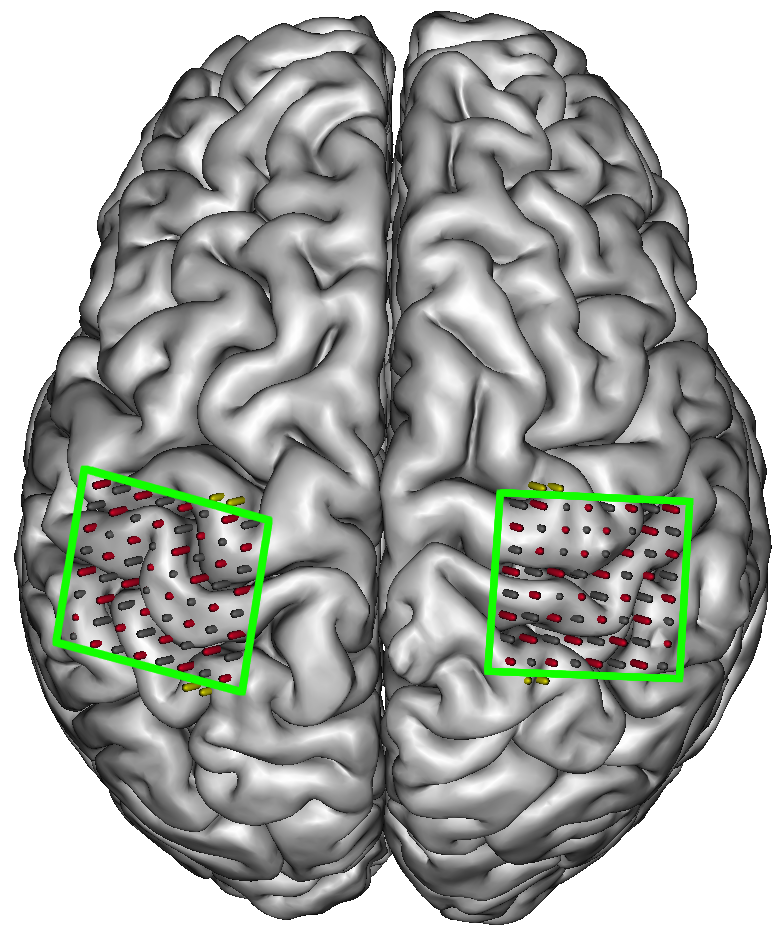}
	\caption{Position of the implants (green) and electrodes (red) on an MRI reconstruction of the patient's brain.} \label{fig:implants-brain}
\end{figure}

Brain signals were recorded by two WIMAGINE® ECoG implants \cite{mestais_wimagine_2015} more than a year after the surgery. Implants were placed over the left and right primary motor and sensory cortex (Figure \ref{fig:implants-brain}) controlling upper limbs. This position aims to provide the best signal for motor imagery decoding \cite{benabid_exoskeleton_2019}. Each implant was composed of an $8 \times 8$ grid of electrodes, but only half of the electrodes, arranged in a chessboard-like manner, were used during the experiment due to the data transfer limit. The signal sampling frequency was 586 Hz and cursor and target positions in the virtual environment were recorded at 10 Hz.

\subsection{Preprocessing and feature extraction}
We extracted time-frequency features from each ECoG channel using continuous complex wavelet transform with 15 Morlet wavelets whose frequencies were regularly chosen between 10 Hz and 150 Hz. The procedure was performed for each one-second-long window $i$ with 90\% overlap. Absolute values of the wavelet coefficients were finally averaged over 0.1s long windows. As a result, we obtained a feature tensor $\underline{\mathbf{X}}_i \in \mathbb{R}^{64 \times 15 \times 10}$ whose dimensions correspond to ECoG channels, frequency bands, and time steps.

\subsection{Loss function}
\begin{figure}
	\centering \includegraphics[width=\columnwidth]{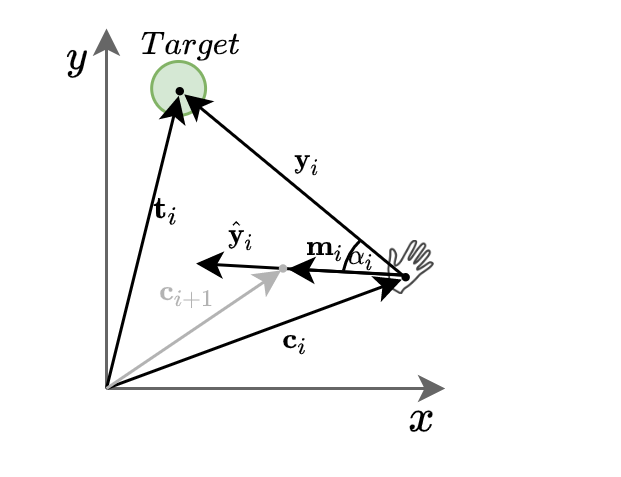}
	\caption{Visualization of movement scheme in 2D. Green circle indicates target position, vectors $\mathbf{c}_i$ and $\mathbf{c}_{i+1}$ are the cursor position vectors for respectively samples $i$ and $i+1$, $\mathbf{t}_i$ is the target position vector, $\mathbf{y}_i$ is the desired trajectory vector for sample $i$, $\mathbf{m}_i$ is the actual move performed by hand based on the prediction $\hat{\mathbf{y}}_i$. $\alpha_i$ is the angle between $\mathbf{y}_i$ and $\hat{\mathbf{y}}_i$.} \label{fig:problem-statement}
\end{figure}

The main problem considered in this study was to predict 3D hand translation from ECoG time-frequency features. At each time step $i$, the desired hand movement $\mathbf{y}_i$ was defined as $\mathbf{t}_i-\mathbf{c}_i$ where $\mathbf{t}_i$ and $\mathbf{c_i}$ respectively correspond to the target and the cursor (point of the avatar hand) position (Figure \ref{fig:problem-statement}). The coordinate system origin was placed in the pelvis of the virtual avatar. Our problem was then to predict $\mathbf{y}_i$ from the feature tensor $\underline{\mathbf{X}}_i$. Hand movement predictions $\hat{\mathbf{y}}_i$ were performed every $\tau = 100$ ms and the cursor moved according to this direction by the vector $\mathbf{m}_i$ until the next prediction. Since predicted movements were rarely fully executed due to hand speed limitation, we compared the predicted and desired vectors with respect to their direction using cosine similarity defined as:
\begin{equation}
\textrm{CS}(\mathbf{y}_i, \hat{\mathbf{y}}_i) = \frac{\mathbf{y}_i \cdot \hat{\mathbf{y}}_i}{\|\mathbf{y}_i\| \cdot \|\hat{\mathbf{y}}_i\|} = \cos{\alpha_i},
\end{equation}
calculating the cosine of the $\alpha_i$ angle between $\mathbf{y}_i$ and $\hat{\mathbf{y}}_i$ vectors. $\textrm{CS} \in [-1, 1]$ is equal to 1 when vectors have exactly the same direction, 0 for orthogonal vectors, and $-1$ for opposite vectors.
We used cosine loss defined as:
\begin{equation}
\textrm{CL}(\mathbf{y}_i, \hat{\mathbf{y}}_i) = 1 - \textrm{CS}(\mathbf{y}_i, \hat{\mathbf{y}}_i)
\end{equation}
as an objective function to train DL models.

\subsection{REW-NPLS}
As a baseline model, we used multilinear models obtained by recursive exponentially weighted N-way partial least squares regression (REW-NPLS) algorithm \cite{eliseyev_recursive_2017}. PLS regression projects both input and response variables to a low-dimensional latent space whose components are designed to provide the highest correlation between input and output variables. This regression method is particularly well-suited in the case of high-dimensional and nonindependent input data. During the experiments, these multilinear models were trained and used to provide real-time control of the avatar to the patient. The latent space dimension was limited to 100 and its optimal dimension was determined every 15 s using recursive validation \cite{eliseyev_recursive_2017}. To obtain the results reported in this study, we retrained the REW-NPLS model in a pseudo-online manner with an update after each 15 s, which simulates the online training. Next, we used this pseudo-online model to compute predictions on the test set.

\subsection{Multilayer perceptron}
Multilayer perceptron (MLP) is a classic method used in machine learning to model complex functions. MLP treats each feature with a set of independent weights to create a representation consisting of neurons excitations that are next processed with nonlinear activation function and finally passed to the next layer. In our experiments, the number of fully connected (FC) layers varied between 1 and 5. Each FC was composed of 50 neurons with ReLu activation. In addition, batch normalization \cite{batch_normalization_ioffe} and dropout \cite{srivastava_dropout} with probability of neuron being zeroed equal $0.5$ were applied between hidden layers (except last). Both methods have a strong regularization effect and are commonly used to increase the generalization of models. This is especially important in the case of small and noisy datasets. The input to MLP was a flattened tensor $\underline{\mathbf{X}}_i$, consisting of 9600 features standardized using Z-score.

\subsection{2D CNN}
One reason convolutional neural networks are widely used in image processing is that they effectively recognize similar patterns in different parts of images. In each layer, several convolutional filters with trainable parameters are shifted over an image to detect patterns and obtain feature maps. This enables CNNs to limit the number of parameters needed to solve complex problems, generalize better, and effectively use information encoded in the local structure of the data and spatial relationships between pixels (for a detailed description of CNN, see explanation by Goodfellow \etal \cite{Goodfellow-et-al-2016}). This kind of architecture can be well-suited for brain signals analysis if its inputs are shaped in a way that enables spatial or temporal pattern detection. Methods proposed further in this section are inspired by architectures used in computer vision \cite{simonyan_very_2015} and EEG signal processing (mainly by Bashivan \etal \cite{bashivan_learning_2016} but also by other methods \cite{schirrmeister_deep_2017, lawhern_eegnet_2018}).

\begin{figure}
	\centering \includegraphics{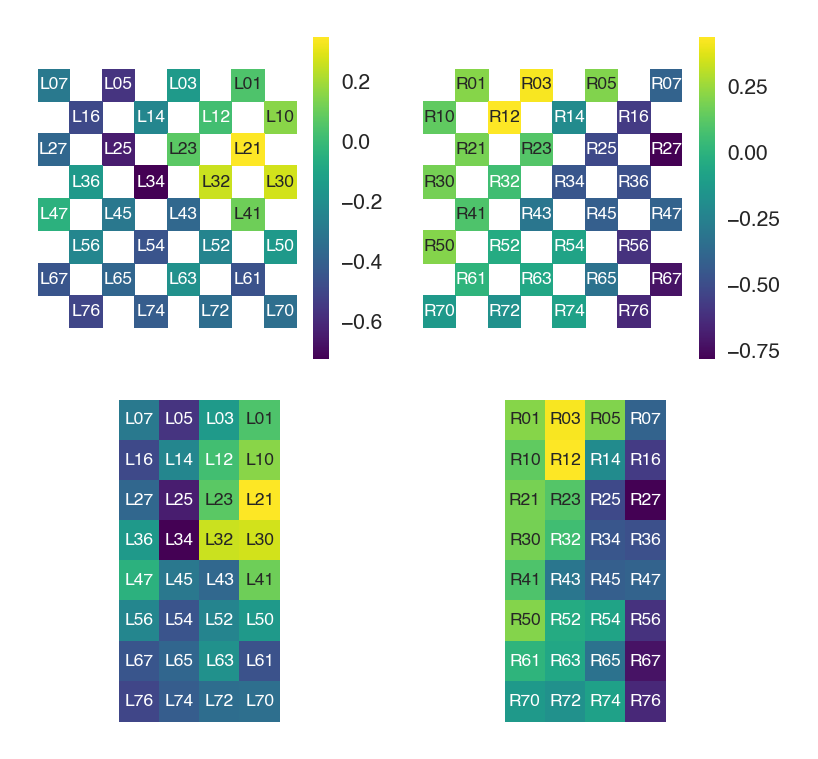}
	\caption{Features from one randomly chosen sample in a frequency band centered at 10 Hz. Values are arranged on a grid approximating actual electrodes' positions for left and right implants, respectively. Top row grids also include electrodes that were not recording signals (white squares). Bottom row plots show electrodes' arrangement after removing missing values. The distance between implants is not preserved.} \label{fig:implants-signal}
\end{figure}

\subsubsection{Input representation and processing} \hfill\\
We propose to preserve the spatial relationship between signals from different electrodes by using an analogous transformation to the one proposed for EEG \cite{bashivan_learning_2016}. Contrary to EEG, ECoG electrodes are distributed on two dense square grids. Hence, to obtain a representation of the actual electrode array, we projected recorded signals onto a grid according to the approximate physical electrodes position presented in Figure \ref{fig:implants-brain}. We created two arrays of 64 electrodes placed on a grid of shape $8 \times 8$ with only half of the electrodes recording signal. We removed unused electrodes and represented each implant with a rectangle of shape $8 \times 4$ merging neighboring columns of electrodes (see Figure \ref{fig:implants-signal}). The introduced distortion of the image does not affect convolution as we preserved the spatial neighborhood of each electrode.

\begin{figure}
	\centering \includegraphics[width=.6\columnwidth]{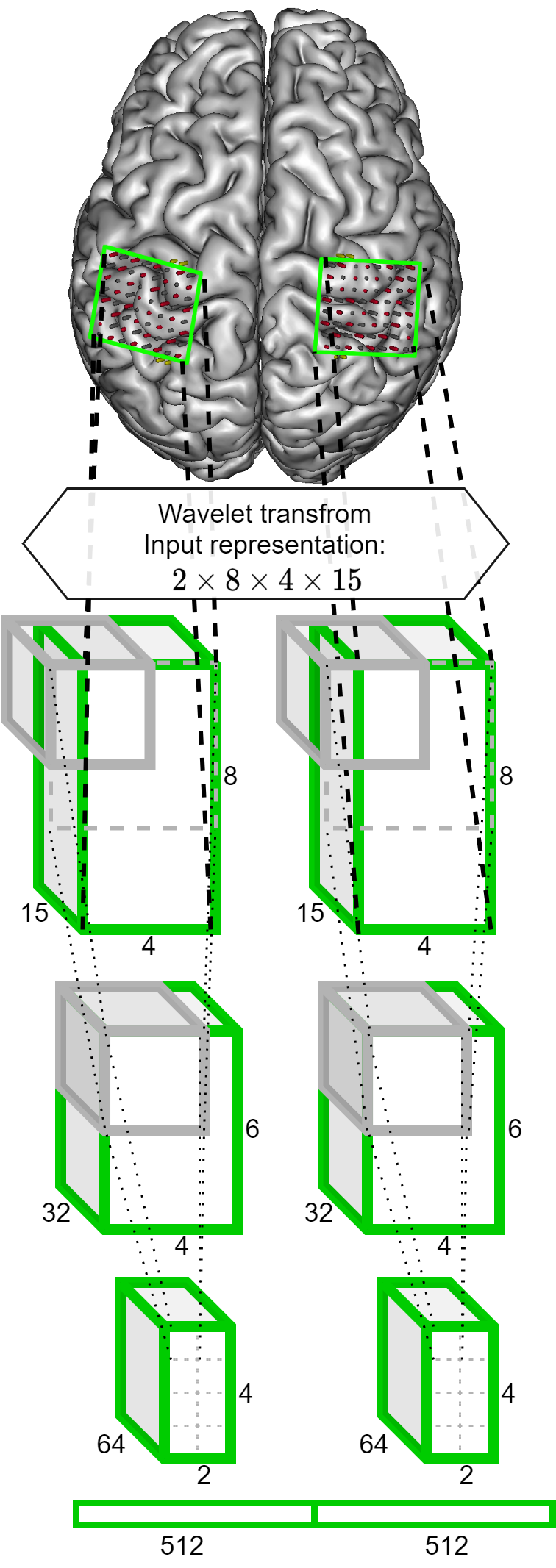}
	\caption{2D CNN processing visualization. Two implants record signals, then time-frequency features are extracted using wavelet transform. Features from each implant (green cuboids) are analyzed using filters (gray cuboid) by consecutive convolutional layers to obtain representation in the form of a vector. We also marked the receptive field (gray dashed rectangle) of the one particular feature from the last layer. Note that this is only a diagram. For simplicity, we visualized a model with two convolutional blocks and we skipped activation, batch normalization, dropout. See detailed specification in Table \ref{tab:cnn2d} and Table \ref{tab:layers}.
	} \label{fig:diagram-cnn-2d}
\end{figure}

Then, the input to all CNN models was a tensor of shape $2 \times 8 \times 4 \times 15 \times 10$, where dimensions correspond to the number of implants, height and width of implants, frequency bands, and time steps, respectively. Features were Z-score standardized using mean and standard deviation per frequency band. Each observation may be interpreted as a time series of consecutive spatio-frequential representations that form two ‘images’ with 15 frequency channels, an analog of the three RGB channels used in computer vision.
Proposed two-dimensional CNN (2D CNN) analyzes each time step independently by performing convolution only in space (two dimensions). The same convolutional kernels were applied separately for each implant (see Figure \ref{fig:diagram-cnn-2d}).

After the final convolutional layer, the features extracted from one second of the signal were flattened and aggregated. We present details of temporal information analysis in sections \ref{sec:cnn2d-fc}-\ref{sec:cnn3d_fc}.

\subsubsection{Convolutional block design} \label{sec:conv-block} \hfill\\
CNNs consist of multiple layers organized in convolutional blocks, usually composed of a convolutional layer and a nonlinear activation function followed by max pooling layer  \cite{Goodfellow-et-al-2016}. Optionally, one can add batch normalization \cite{batch_normalization_ioffe} and dropout \cite{srivastava_dropout}. The optimal structure of convolutional block depends on the characteristic of the problem, e.g., size and type of data, signal, and problem characteristic. Further in this section, we describe the convolutional block design choices that we made in the study.

\paragraph{Batch normalization and dropout} \hfill\\
We decided to use both dropout and batch normalization to achieve a strong regularization effect, as our dataset suffers from a small number of samples, low signal to noise ratio, and uncertain labels. We decided to include batch normalization between the activation and the dropout layer, which might be more effective in the case of ReLU activation \cite{chen_rethinking_2019}. Merged batch normalization and dropout can be interpreted as an Independent-Component (IC) layer, reducing mutual information between neurons in the input \cite{chen_rethinking_2019}. The batch normalization layer was included in all convolutional blocks except last.

\paragraph{No-padding} \hfill\\
Another design choice that we propose to use is to remove the max pooling layer. Instead, we used no-padding option (removing padding around grid edges) to reduce the size of the ‘images’ (see padding sizes in Table \ref{tab:layers}). We considered this variant as a reasonable choice because our ‘images' (size $8 \times 4$) are much smaller than typical computer vision images (e.g. $224 \times 224 \times 3$ for ResNet \cite{he_deep_2015}) and spatio-frequential EEG ‘images' ($32 \times 32 \times 3$ analyzed by Bashivan \etal \cite{bashivan_learning_2016}). Reducing the size of our images too much could have resulted in the loss of useful information. A convolution operation (kernel size $3 \times 3$ and stride equal to 1) without padding reduces an image size of 2 pixels along each dimension, whereas a max pooling operation (kernel size $2 \times 2$, stride 2) halves its dimension.

\paragraph{Activation function} \hfill\\
Following the results presented by Schirrmeister \etal \cite{schirrmeister_deep_2017} and Lawhern \etal \cite{lawhern_eegnet_2018} regarding activation function choice, we considered two potential candidates---ReLU, used by Bashivan \etal \cite{bashivan_learning_2016}, and exponential linear unit (ELU) \cite{clevert2016fast} that was proven to be more effective in the case of EEG processing \cite{schirrmeister_deep_2017, lawhern_eegnet_2018}.\\

Finally, we used convolutional blocks consisting of: convolutional layer \verb|->| ReLU \verb|->| batch normalization \verb|->| dropout. All convolutional layers use typical parameters: kernel size equal to $3 \times 3$, stride equal to 1, and variable size of zero padding (denoted in Table \ref{tab:layers}). For dropout we used probability of zeroing whole channel equal $0.5$.

\begin{table}[]
\caption{Architecture specification of 2D CNN model with two convolutional blocks.} \label{tab:cnn2d}
\centering
\resizebox{\columnwidth}{!}{%
\begin{tabular}{@{}lllll@{}}
\br
Block                                   & \multicolumn{1}{l}{Layer} & Filters & Padding & Output shape                     \\ \br
\multicolumn{1}{c}{}                   & Input                     &         &         & $2 \times 8 \times 4 \times 15$  \\ \mr
\multicolumn{1}{c}{\multirow{4}{*}{1}} & Convolution               & 32      & (0, 1)  & $2 \times 6 \times 4 \times 32$  \\
\multicolumn{1}{c}{}                   & ReLU                      &         &         & $2 \times 6 \times 4 \times 32$  \\
\multicolumn{1}{c}{}                   & Batch normalization       &         &         & $2 \times 6 \times 4 \times 32$  \\
\multicolumn{1}{c}{}                   & Dropout                   &         &         & $2 \times 6 \times 4 \times 32$  \\ \mr
\multicolumn{1}{c}{\multirow{4}{*}{2}} & Convolution               & 64      & (0, 1)  & $2 \times 4 \times 2 \times 64$  \\
\multicolumn{1}{c}{}                   & ReLU                      &         &         & $2 \times 4 \times 2 \times 64$  \\
\multicolumn{1}{c}{}                   & Dropout                   &         &         & $2 \times 4 \times 2 \times 64$  \\ \br
\end{tabular}
}
\end{table}

\begin{table*}[]
	\caption{Description of the 2D CNN architectures with different number of layers. The parameters of convolutional blocks are indicated in the form 'conv(padding height, padding width)-$<$number of channels$>$'. The last row presents the number of extracted features.} \label{tab:layers}
	\centering
		\begin{tabular}{|c|c|c|c|c|}
			\hline
			\multicolumn{5}{|c|}{\textbf{2D CNN}}                                            \\ \hline
			1 layer       & 2 layers      & 3 layers       & 4 layers       & 5 layers       \\ \hline
			\multicolumn{5}{|c|}{Input size:  $2 \times 8 \times 4 \times 15$}               \\ \hline
			conv(0, 0)-32 & conv(0, 1)-32 & conv(0, 1)-32  & conv(0, 1)-32  & conv(0, 1)-32  \\ \hline
			&               &                & conv(1, 1)-32  & conv(1, 1)-32  \\ \hline
			& conv(0, 0)-64 & conv(0, 1)-64  & conv(0, 1)-64  & conv(0, 1)-64  \\ \hline
			&               &                &                & conv(1, 1)-64  \\ \hline
			&               & conv(0, 0)-128 & conv(0, 0)-128 & conv(0, 0)-128 \\ \hline
			\multicolumn{5}{|c|}{Flatten}                                                    \\ \hline
			768           & 1024          & 1024           & 1024           & 1024           \\ \hline
		\end{tabular}
\end{table*}

\subsubsection{Number of blocks} \hfill\\
Another important architecture choice is the number of convolutional blocks. In computer vision, typical models consist of dozens of convolutional layers (e.g. ResNet \cite{he_deep_2015} with more than 1000 convolutional layers, VGG \cite{simonyan_very_2015} with up to 19 layers, Inception-v3 \cite{szegedy_rethinking_2016} with 48 layers). However, methods proposed for brain signals analysis used a significantly lower number of layers (e.g., ShallowConvNet and DeepConvNet \cite{schirrmeister_deep_2017} with two and five convolutional layers respectively, EEGNet \cite{lawhern_eegnet_2018} with three layers, Bashivan \etal \cite{bashivan_learning_2016} proposed architectures with up to seven convolutional layers). To investigate this problem, the number of convolutional blocks in the 2D CNN was set between 1 and 5 (Table \ref{tab:layers}). The optimal model depth was selected based on the session-wise cross-validated CS obtained with the simplest proposed architecture: CNN2D+FC (see Section \ref{sec:cnn2d-fc}). This depth value was then used for more complex approaches.

\begin{figure*}
	\centering \includegraphics[width=\textwidth]{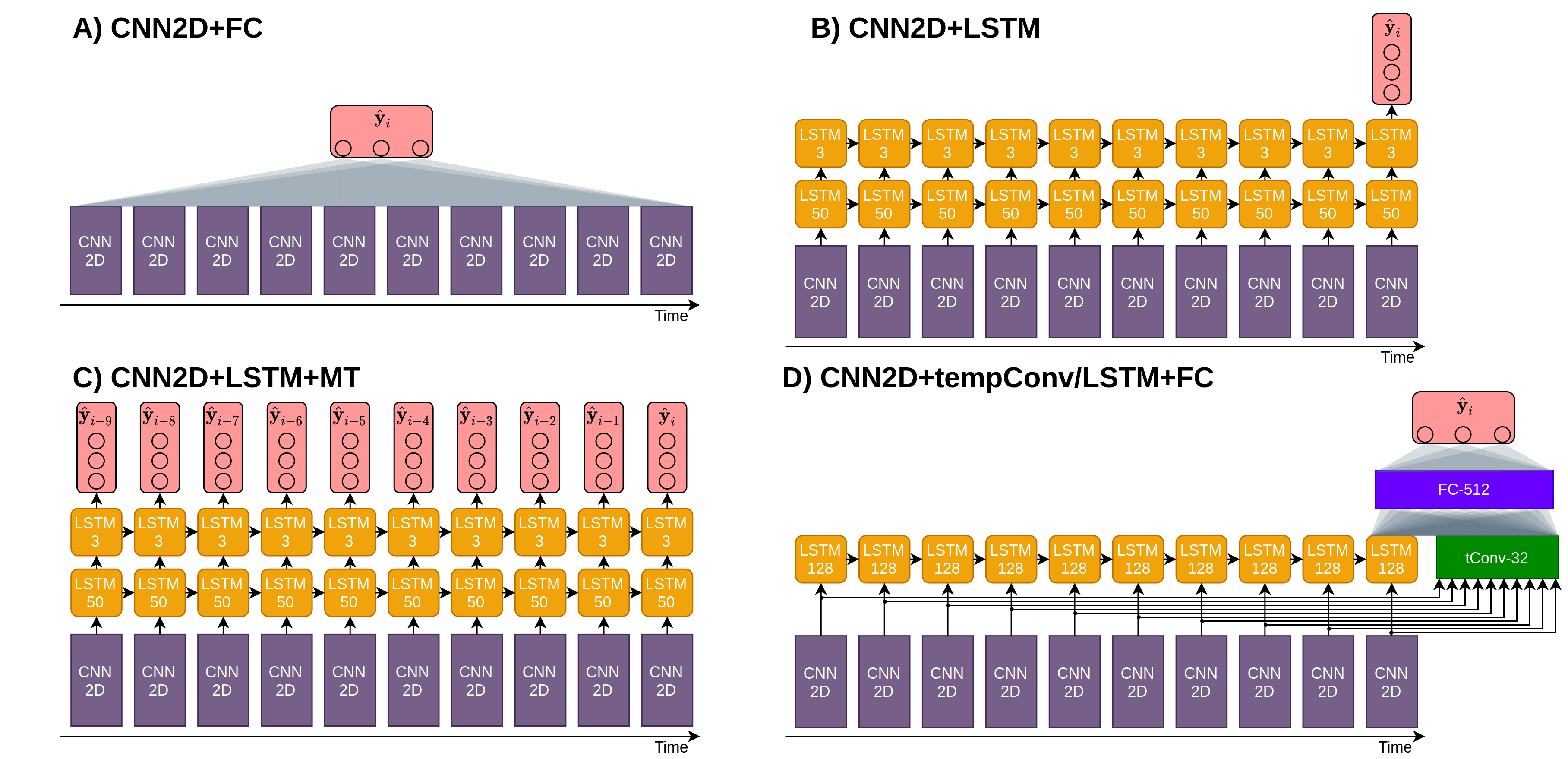}
	\caption{Four ways of temporal aggregation are presented for architectures based on the 2D CNN model. Numbers after the name of the layer denote hidden state size in the case of LSTM, filter number in the case of tConv, and the number of neurons in the case of FC. Visualization inspired by figures presented by Bashivan \etal \cite{bashivan_learning_2016}.} 
	\label{fig:temporal} 
\end{figure*}

\subsection{CNN2D+FC} \label{sec:cnn2d-fc} \hfill\\
We proposed several architectures to aggregate temporal information from representation extracted by the 2D CNN model. In the most straightforward approach, the features extracted by 2D CNN at each time step were concatenated and given as an input to an FC layer composed of three neurons (Figure \ref{fig:temporal}.A). It provided three outputs that correspond to $\hat{\mathbf{y}}_i$ components. Further, we will refer to this approach as CNN2D+FC.

\subsection{CNN2D+LSTM} \hfill\\
Long short-term memory network (LSTM) \cite{hochreiter_long_1997} is a type of recurrent neural network that can efficiently analyze long temporal relationships in the data. As in typical RNNs, a module called cell is applied to each time step. LSTM cell carries relevant information through time and decides what to forget and what to store based on the current input \cite{colah_lstm}. In order to analyze temporal information, we stacked two LSTM layers at the top of 2D CNN (see Figure \ref{fig:temporal}.B). At each time step, the first LSTM cell with a hidden state of size 50 was provided with the flattened output features of the 2D CNN. The second LSTM layer had three neurons that were used to provide three output coordinates for $\hat{\mathbf{y}}_i$ components. Stacking two LSTM layers enabled the network to model complex temporal relationships and efficiently incorporate information from the last second of the signal. It was already proven effective in a variety of tasks including ECoG decoding \cite{xie_decoding_2018,du_decoding_2018,elango_sequence_2017}.

\subsubsection{CNN2D+LSTM+MT} \hfill\\
As an extension of CNN2D+LSTM, we also tested a modification, referred to readers as multiple trajectory prediction (MT), in which each LSTM output from the last layer is used for network training (see Figure \ref{fig:temporal}.C). It means that $\hat{\mathbf{y}}_i$ and $\mathbf{y}_i$ were compared for ten consecutive time steps to compute the model error. This was possible because desired trajectory vector $\mathbf{y}_i$ was registered every 100 ms. Thus, LSTM may utilize the relation between close desired trajectory to create a more general representation of the system state. As a consequence, the loss function was modified as follows:
\begin{equation}
\mathrm{CL_{MT}}(\mathbf{y}_i, \hat{\mathbf{y}}_i) = \sum_{j=i-N-1}^{i} \mathrm{CL}(\mathbf{y}_j, \hat{\mathbf{y}}_j),
\end{equation}
where $N$ is the number of $\mathbf{y_i}$ recorded during a one-second-long window. In our case, $N$ was equal to 10 because the desired trajectory was recorded at 10 Hz. 

\subsection{CNN2D+tConv/LSTM+FC} \hfill\\
To compare our architectures with the state-of-the-art, the best temporal aggregation variant proposed by \cite{bashivan_learning_2016} (variant D) was included in our analysis.  For consistency, we will refer to this approach as CNN2D+tConv/LSTM+FC. Bashivan \etal \cite{bashivan_learning_2016} used one LSTM layer with a hidden state of size 128 to analyze temporal information in parallel with temporal convolution. The temporal convolution (tConv) consisted of 32 filters of size 3 that were shifted over features extracted by 2D CNN in the time domain. It enables recognizing time-invariant patterns in the data (similar patterns that occur at different moments). Finally, the output of tConv and the LSTM cell from the last time step were concatenated and fed into an FC layer with 512 neurons followed by an FC with 3 neurons predicting $\hat{\mathbf{y}}_i$  (see Figure \ref{fig:temporal}.D). A dropout layer was added before each FC layer.

\subsection{CNN3D+FC} \label{sec:cnn3d_fc} \hfill\\
We tested one more approach that can analyze spatial and temporal patterns at the same time at all levels of the data processing. Inspired by three-dimensional CNNs (3D CNNs) \cite{Ji20133DCN} used to recognize human actions on videos, we propose to extend the 2D CNN model and to perform convolution not only in space but also in time (see Table \ref{tab:cnn3d}). These architectures have the advantage of propagating temporal information from the first to the last layer. Similar to 2D CNN, an FC was used as the output layer and the same convolutional parameters were chosen (except for convolution performed in time).

\begin{table}[]
\caption{Architecture specification of 3D CNN model with two convolutional blocks.} \label{tab:cnn3d}
\centering
\resizebox{\columnwidth}{!}{%
\begin{tabular}{@{}lllll@{}}
\br
Block                                   & \multicolumn{1}{l}{Layer} & Filters & Padding & Output shape                     \\ \br
\multicolumn{1}{c}{}                   & Input                     &         &         & $2 \times 8 \times 4 \times 15 \times 10$  \\ \mr
\multicolumn{1}{c}{\multirow{4}{*}{1}} & Convolution               & 32      & (0, 1, 0)  & $2 \times 6 \times 4 \times 32 \times 8$  \\
\multicolumn{1}{c}{}                   & ReLU                      &         &         & $2 \times 6 \times 4 \times 32 \times 8$  \\
\multicolumn{1}{c}{}                   & Batch normalization       &         &         & $2 \times 6 \times 4 \times 32 \times 8$  \\
\multicolumn{1}{c}{}                   & Dropout                   &         &         & $2 \times 6 \times 4 \times 32 \times 8$  \\ \mr
\multicolumn{1}{c}{\multirow{4}{*}{2}} & Convolution               & 64      & (0, 0, 0)  & $2 \times 4 \times 2 \times 64 \times 6$  \\
\multicolumn{1}{c}{}                   & ReLU                      &         &         & $2 \times 4 \times 2 \times 64 \times 6$  \\
\multicolumn{1}{c}{}                   & Dropout                   &         &         & $2 \times 4 \times 2 \times 64 \times 6$  \\ \br
\end{tabular}
}

\end{table}

\subsection{Sensitivity analysis}
To analyze model behavior and identify important features, we computed the gradient of the model's outputs with respect to the inputs on the test datasets---network jacobian $\underline{\mathbf{J}} \in \mathbb{R}^{N \times 2 \times 8 \times 4 \times 15 \times 10 \times 3}$ representing the sensitivity of each network output to the inputs. For each input feature, the higher the absolute gradient value, the stronger the influence on the prediction. For visualization, the sensitivity of the three outputs was averaged. We analyzed feature importance in three domains: spatial with projection on two implants, frequential within 15 frequency bands from 10 Hz to 150 Hz, and temporal within 10 time steps containing 1 second of the signal.

\subsection{Optimization, hyperparameters, and evaluation}
We selected optimal hyperparameters values (initial learning rate, weight decay, and batch size) with the Tree of Parzen Estimators (TPE) \cite{bergstra_algorithms_2011} algorithm from hyperopt package. The selection was performed on the left hand calibration dataset using session-wise 6-fold cross-validation. Hyperparameters values selected after 200 iterations can be found in the Appendix in Table \ref{tab:hyperparams-value}. 

Models were trained for 60 epochs with early stopping (to limit overfitting) with the patience of 20 epochs without any loss function improvement on the validation dataset. The learning rate was changed using cosine annealing \cite{loshchilov2017sgdr}. The last 10 \% of the calibration dataset was used for validation and the rest was used for training. We discarded the random cross-validation scheme as the high correlation between neighboring samples in time would have biased performance on the validation datasets. The model which achieved the best score on the validation dataset was retained. It was used to compute average CS on the test dataset (Table \ref{tab:cs-overall}).

To limit the influence of network weights initialization and the optimization process on the results, each model was trained five times. The mean and standard deviation of the performance indicator was computed. We used a T-test for independent samples to assess the statistical significance of the difference in performance between architectures.

We used PyTorch \cite{NEURIPS2019} and skorch \cite{skorch} for DL models training and evaluation, MATLAB \cite{MATLAB2017} for multilinear models training and evaluation, Seaborn \cite{Waskom2021} and Matplotlib \cite{Hunter2007} for data visualization, and Pandas \cite{pandas} for results analysis.

\section{Results}
Our analysis started with determining the optimal number of layers in the 2D CNN model and the MLPs. Next, we compared all proposed methods in terms of CS on the left and right hand datasets. Then, a detailed comparison of the best DL-based architecture and the state-of-the-art multilinear model is presented. Finally, we analyzed the influence of particular design choices on DL models' performance.

\subsection{Number of layers}
\begin{figure}[ht]
	\centering \includegraphics{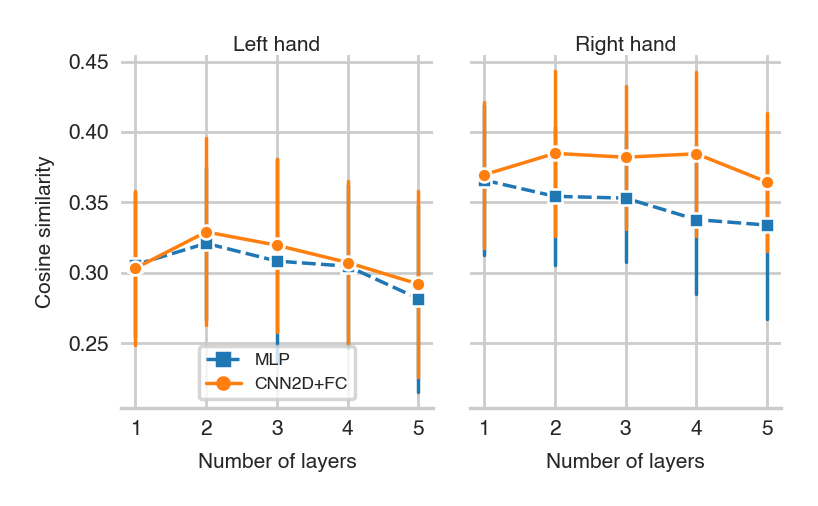}
	\caption{Influence of the number of layers on model performance. Error bars denote the standard deviation of five runs.} \label{fig:number-of-layers}
\end{figure}

The influence of the number of layers on the calibration cross-validated accuracy of MLP and CNN2D+FC is presented in Figure \ref{fig:number-of-layers}. In the LH case, the best CS was obtained with two layers for both architectures. For the RH, the best CS was obtained with one layer for MLP and again with two layers for CNN. We can observe a decrease in performance when adding more layers starting from 2 (LH) or even 1 (MLP and RH). CNN architecture performed more stable on the RH dataset and decreased accuracy was observed for more than four layers. Based on these results, the number of blocks in 2D CNN and the number of hidden FC layers in MLP were chosen to be two as this choice maximized the average CS over LH and RH results. It simplified further analysis and limited computation times without decreasing the performance.

\subsection{Overall model performance}

\begin{figure*}[ht]
	\centering \includegraphics{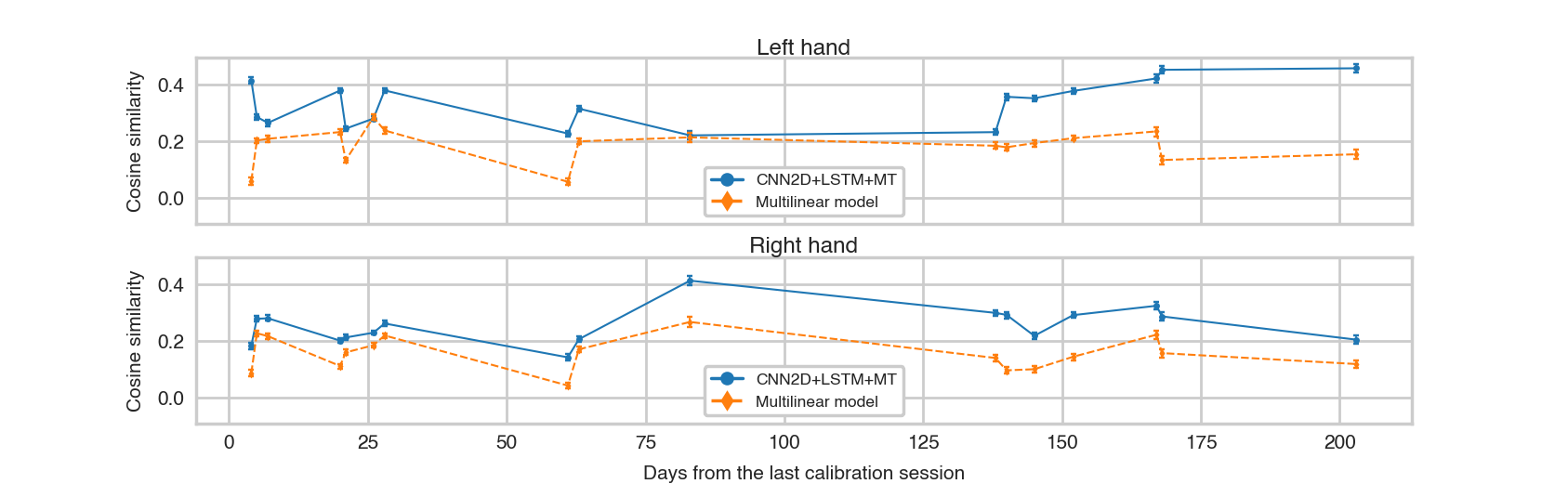}
	\caption{The course of cosine similarity of CNN2D+LSTM+MT and multilinear model in time. Error bars denote 95\% mean confidence interval.} \label{fig:cs-sessions}
\end{figure*}

The CS of all evaluated approaches on both test datasets is given in Table \ref{tab:cs-overall}. The CNN2D+LSTM+MT model achieved the best average performance across left and right hand datasets with a CS of 0.302 for the left hand and 0.249 for the right hand, corresponding to 60\% and 59\% of CS improvement relative to the multilinear model. Comparisons using the T-test showed significant differences ($p \textless{0.05}$) for both hands between CNN2D+LSTM+MT and CNN2D+LSTM, MLP, and multilinear model. Additionally, in the case of the RH dataset, CNN2D+LSTM+MT was also significantly better than CNN3D+FC and CNN2D+tConv/LSTM+FC. All proposed DL methods performed better than the multilinear model. Generally, all models decoded LH movements more accurately than RH movements.

\begin{table}[ht]
	\caption{Cosine similarity for each proposed method for left and right hand datasets. Asterisks denote models which had significantly different cosine similarity in comparison to CNN2D+LSTM+MT.} \label{tab:cs-overall}
	\centering
	\lineup
	\resizebox{\columnwidth}{!}{
		\begin{tabular}{lll}
			\br
			\textbf{}               & \multicolumn{2}{c}{\textbf{Cosine similarity}}          \\
			& Left hand                  & Right hand                 \\ \mr
			CNN2D+LSTM+MT       & $0.302 \pm  0.017$ & $\mathbf{0.249 \pm 0.008}$ \\
			CNN2D+FC            & $0.296 \pm 0.015$           & $0.237 \pm 0.011$         \\
			CNN2D+tConv/LSTM+FC & $\mathbf{0.306 \pm 0.017}$  & $0.223 \pm 0.012$*          \\
			CNN3D+FC            & $0.294 \pm 0.016$           & $0.226 \pm 0.011$*         \\
			CNN2D+LSTM          & $0.264 \pm 0.015$*          & $0.222 \pm 0.011$*         \\
			MLP                 & $0.229 \pm 0.01$*          & $0.205 \pm 0.011$*         \\
			Multilinear model   & 0.189*                      & 0.157*                     \\ \br
		\end{tabular}
	}
\end{table}

\subsection{Detailed performance comparison}
To understand why CNN2D+LSTM+MT outperformed the multilinear models, we performed a detailed analysis of the accuracy of both methods. The CNN2D+LSTM+MT model used for comparison was randomly selected from the five models trained for that study. Then, the models' performance was compared over time, depending on the distance to the targets and depending on the desired trajectory direction.

Firstly, we compared the stability of the performance over time. The average CS on each day of the experiment is given in Figure \ref{fig:cs-sessions}. The DL model obtained a higher CS for a majority of points (15/17) in the case of the LH dataset and all in the RH dataset. For the LH and in comparison to the multilinear model, the CS obtained by the DL models was better (except the 26th day) until the 83rd day, similar on days 83rd and 138th, and even much better until the end of the test set. The improvement is more uniform for the right hand, with an increase starting from the 83rd day after the last calibration session. The CS of CNN2D+LSTM+MT varies similarly to the multilinear REW-NPLS model in the case of the RH dataset. CS was better than zero for all recording days for both models.

\begin{figure}[ht]
	\centering \includegraphics{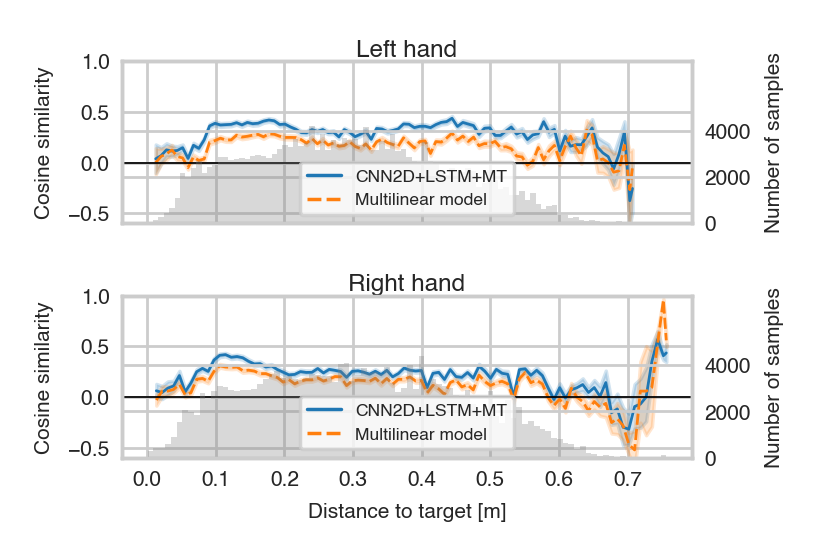}
	\caption{Cosine similarity as a function of the distance between cursor and target. The gray histograms represent the number of samples in the test dataset. Error bands denote 95\% mean confidence interval.} \label{fig:cs-distance-hist}
\end{figure}

The accuracy depending on the distance between the hand cursor and the target is shown for both models in Figure \ref{fig:cs-distance-hist}. The DL model achieved a higher CS than the multilinear model for both hands in almost the whole range of distance (\textgreater{}90\%). For both hands and models, a drop in performance occurred when the distance to the target was inferior to 10 cm. The CS variance increased strongly when this distance was superior to 60 cm. There are significantly fewer observations for the edge values of distance to the target, so the performance estimate is noisier.

\begin{figure}[ht]
	\centering \includegraphics{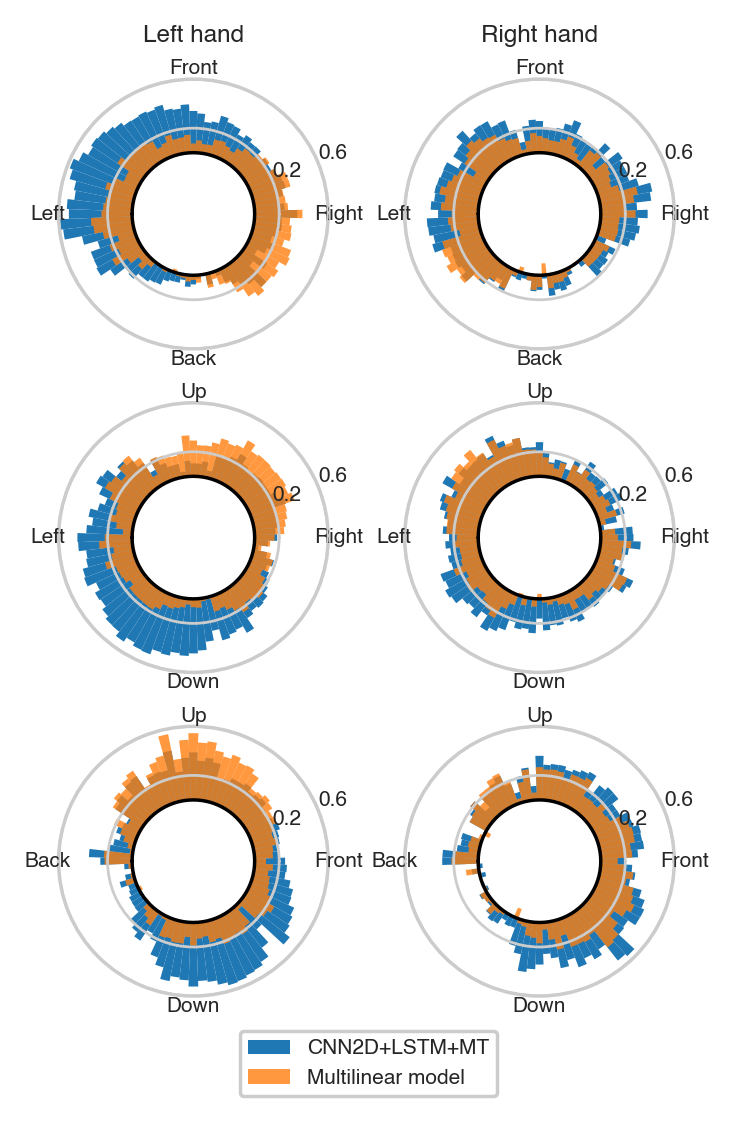}
	\caption{Cosine similarity directional performance of multilinear and CNN2D+LSTM+MT models projected on 2D planes.} \label{fig:directional-cs}
\end{figure}

To determine if the predictions were more accurate in a given direction, we plotted CS on 2D planes (Figure \ref{fig:directional-cs}). CS has a non-uniform distribution---observations for which the patient was asked to move hand backward have a lower CS. CNN2D+LSTM+MT and multilinear model performed better in different directions. There is no wide angle in which performance is below 0.

\subsection{Sensitivity analysis}
\begin{figure}
	\centering \includegraphics[width=\columnwidth]{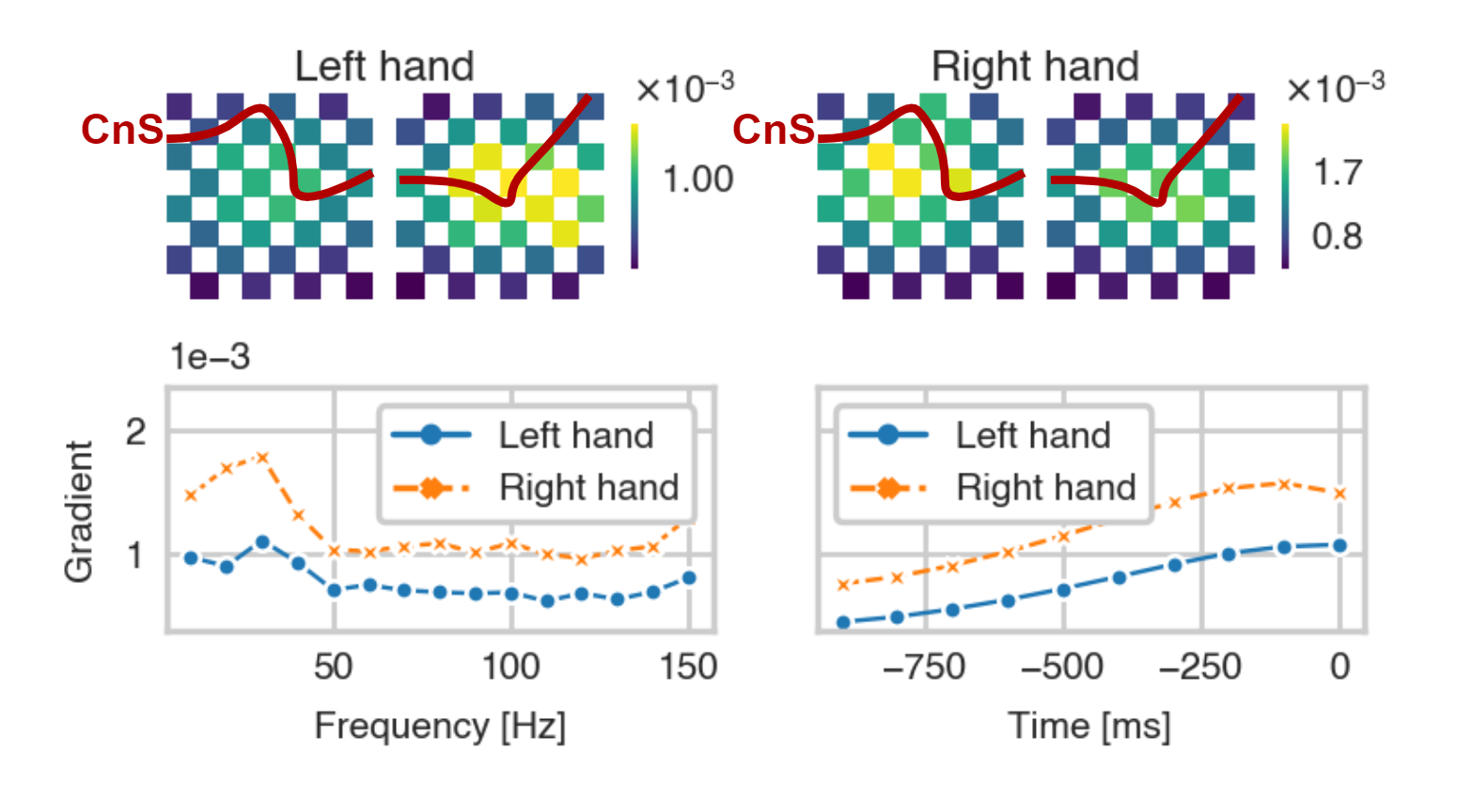}
	\caption{Average sensitivity of the CNN2D+LSTM+MT outputs on the test dataset. Absolute gradient values are plotted as a function of electrode positions (top plots), frequency (left bottom plot), and time (right bottom plot). CnS corresponds to the red line on the heatmaps.}
 
	\label{fig:gradient-viz-sulcus}
\end{figure}

To determine the inputs that had the strongest influence on the prediction, we visualized the average sensitivity of the DL model on the test set (Figure \ref{fig:gradient-viz-sulcus}). We put together spatial feature importance maps and the approximate shape of the central sulcus (CnS). CnS separates the primary motor cortex and the primary somatosensory cortex. For each hand, we can see that the DL model outputs were more sensitive to changes in ECoG features from the contralateral implant.
The highest sensitivity can be observed for the electrodes that are close to the CnS. For the LH movements decoding, the most important electrodes are located in the center of the implant surrounding the CnS of the right cortex, while in the case of the RH, they are posterior to the CnS of the left cortex. The lowest absolute gradient values can be found at the top and bottom edges of the implants.

The most important features in the frequency domain correspond to 20 Hz and 30 Hz (Beta rhythm). An increase in importance can also be observed in frequencies higher than 130 Hz.

In the time domain, the importance of input features increases when getting closer to the current time step, with a decrease observed for the features computed with the last 100 ms of the ECoG signal for the RH movements prediction.

\subsection{Multiple trajectory influence}
To understand better the influence of the MT variant on the performance, we modified $CL_{MT}$ loss function. At time step $j \in [i-N-1, i]$ instead of comparing LSTM output $\hat{\mathbf{y}}_j$ to the corresponding desired trajectory $\mathbf{y}_j$, it was compared to the desired trajectory at the last time step $\mathbf{y}_i$ (the one that is used when only one time step is predicted).  Finally, the modified loss function was defined as:
\begin{equation}
\mathrm{CL_{\mathcal{MT}}}(\mathbf{y}_i, \hat{\mathbf{y}}_i) = \sum_{j=i-N-1}^{i} \mathrm{CL}(\mathbf{y}_i, \hat{\mathbf{y}}_j),
\end{equation}
This enabled us to isolate the influence of the information about desired trajectory variation on the performance from other factors as changes in the optimization process due to providing explicit gradient to the LSTM cell at each time step.
Models trained with $\mathrm{CL_{\mathcal{MT}}}$ loss function obtained cosine similarity of $0.285 \pm 0.012$ and $0.224 \pm 0.0077$ for the left and right hand respectively. This result is not as good as in the case of standard MT modification (LH: $0.302 \pm  0.009$, RH: $0.246 \pm 0.011$).

\subsection{Influence of convolutional block design}
\begin{figure*}
	\centering \includegraphics{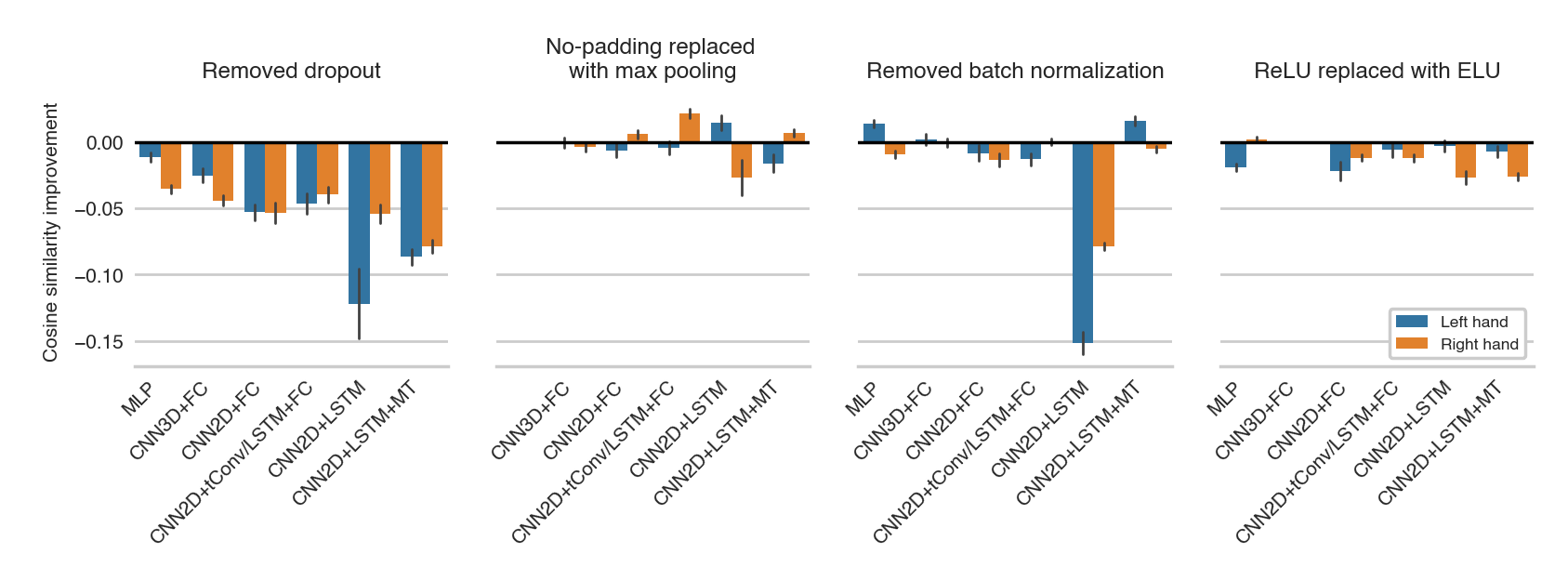}
	\caption{Influence of the architecture choices on the model performance. Models were trained with and without modification for each architecture choice (dropout, no-padding, batch normalization). The difference between the two obtained models was computed to evaluate the influence of particular modifications on the model performance. The lower the cosine similarity difference is, the bigger deterioration is caused by removing the proposed modification from the model.} \label{fig:architecture-conv-blocks}
\end{figure*}
The convolution blocks had the following chosen structure: dropout (p=0.5), batch normalization, ReLU activation function, and no-padding. To assess the influence of each particular design choice in the convolutional block, we individually removed dropout, batch normalization, replaced no-padding with max pooling, and replaced ReLU with ELU. We compared CS obtained over the test set by each DL model with and without the particular design choices. Results are presented in Figure \ref{fig:architecture-conv-blocks}. It enabled us to separately estimate the deterioration/improvement related to each design choice. When CS values are negative, it means that not applying the design choice decreased the accuracy. The dropout layer brought the biggest improvement overall for both hands and all CNN-based models except CNN2D+LSTM where batch normalization had a stronger impact. We observed much smaller improvements in the case of no-padding, batch normalization, and activation function.

\begin{figure}[ht]
	\centering \includegraphics{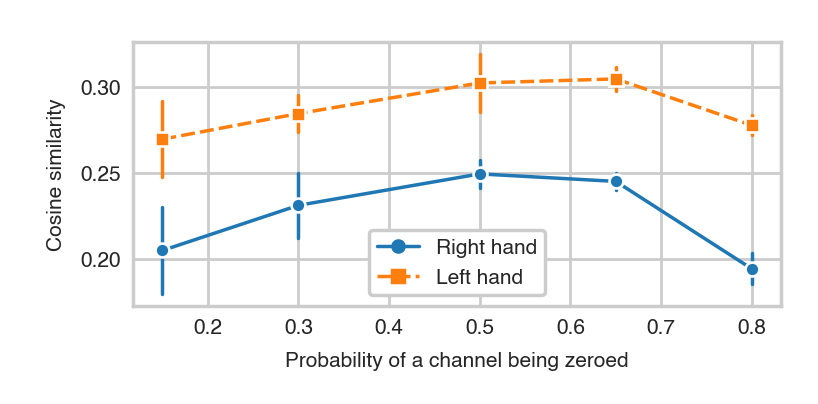}
	\caption{Cosine similarity obtained with CNN2D+LSTM+MT model with two convolutional blocks as a function of dropout levels.} \label{fig:dropout-level}
\end{figure}

As the biggest improvement in CS was due to the dropout, we investigated its optimal value. The CNN2D+LSTM+MT model was trained with different probabilities of zeroing a channel. CS for each hand for different dropout values is given in Figure \ref{fig:dropout-level}. Results look similar for both hands: optimal dropout values were 0.5 and 0.65. High (0.8) or low ($\leq0.15$) dropout value strongly decreased the network accuracy.

\section{Discussion}
We proposed several DL-based methods to predict 3D hand translation. Our offline results show that both MLP and CNN-based models outperform the multilinear model proposed in \cite{eliseyev_recursive_2017}. A significant improvement was obtained with CNN-based methods in comparison to both MLP and multilinear models.
At the same time, CNN2D+LSTM+MT uniformly improved cosine similarity, independently to the distance to the target, and provided better performance than multilinear models in the majority of directions.

We demonstrated that CNNs are a reasonable choice to analyze time-frequency ECoG features and may be used to predict complex 3D hand translation. All the presented results were computed offline, without any interaction between model and patient. This allowed us to separate decoder performance and patient adaptation influence and finally compare models in an isolated environment. Our dataset was recorded in a closed-loop experiment, which includes patient's corrections after erroneous movements and control feedback that influence the user's brain activity. Thus, our estimate of models' performance should be closer to the one obtained during online control compared to open-loop experiments. However, training on previously recorded trajectories may lead to overfitting and accommodation to the model used during the experiment and the model's errors. This can impact the online results, but it is impossible to measure those factors' influence on the results computed in an offline study.

In our dataset, we did not have access to real hand movements. We analyzed only movements imagined by a tetraplegic patient. Compared to the open-loop experiment with actual movement recorded, our target variable can be distorted and influenced by the patient's attention level, tiredness, or inexactness of imagination. Moreover, in a closed-loop, a patient has to correct the erroneous movements, which complicates the trajectory. As a result, we train models on a distorted and noisy dataset. It also complicates model evaluation and makes the problem more challenging than in overt movements.

\subsection*{Generalization of the results}
This study included only one participant. It is then impossible to generalize these results to other patients. However, the proposed models were tested on two different datasets. CNN2D+LSTM+MT model achieved a similar improvement in comparison to the state-of-the-art multilinear model and MLP for both hands. The sensitivity analysis demonstrated that the imagination patterns produced for each hand were different. Thus, it showed the DL-based models' capacity to determine the most relevant features on two datasets. Nevertheless, our hypotheses must be verified with additional tests (including online evaluation) of proposed models on a bigger group of patients. This is planned in the clinical trial as the next steps after enrollment of new patients.

\subsection*{Online training}
REW-NPLS enables incremental training of multilinear models from newly recorded data chunks. This is particularly important in a closed-loop experiment since it enables co-adaptation between the patient and the models. Such kind of training may be hard to use in the case of DL. Backpropagation may fail to find a reasonable solution \cite{ijcai2018-369}, especially when provided with small chunks (e.g., 150 samples corresponding to 15 seconds of signal in the case of REW-NPLS) that are likely to be biased towards an overrepresented direction. Such circumstances may lead to a drastic decrease in DL model performance. To train DL models incrementally over the ECoG data stream, one may increase the data chunk size or keep the whole or part of the dataset in the memory and mix it with new data. Schwemmer \etal \cite{schwemmer_meeting_2018} updated a pretrained model using a part of their training dataset combined with newly recorded data. One can also utilize more sophisticated methods like hedge backpropagation \cite{ijcai2018-369}. An alternative solution could be keeping all the data in memory and retraining from scratch each time a new data chunk is available. This solution is memory inefficient and may not be possible to perform in real-time as training from scratch may take more than 2 minutes (for CNN2D+LSTM+MT and around 40 minutes of signal). More experiments must be conducted to study the possibility of training DL models incrementally. Nevertheless, our study shows that a standard optimization of the DL-based models from data recorded during multilinear model incremental training enables obtaining significantly better models.

\subsection*{2D CNN}
Methods processing data with 2D CNN obtained higher cosine similarity compared to solutions based on flattened feature vector. Input representation for 2D CNN enabled exploitation of local correlation between electrodes in contrast to MLP and multilinear model. 2D CNN model with several convolutional blocks also has significantly fewer trainable parameters than MLP. In 2D CNN architecture, the data processing is separated for each implant. We also considered a scenario in which the features from both implants would be concatenated along the width dimension as if there was no space between the inner edges of the implants. However, we decided to keep a separate data processing for each implant to avoid introducing an additional spatial distortion that would have broken the spatial consistency between neighboring electrodes. Then, the same set of weights was used for both implants. This enabled us to halve the number of trainable parameters and to learn features that generalize across implants. Nevertheless, MI imagery patterns recorded by each ECoG implant are different since they are not positioned in the same cerebral hemisphere. From neuroscience, we also know that most of the brain activity correlated to unilateral hand MI/movements occurs in the contralateral brain hemisphere. However, brain activity correlated to this kind of movement can also be found in the ipsilateral motor cortex \cite{Bundy10042}. Due to the weight sharing property of CNNs, the network is oriented towards extracting low-level features that are implant invariant. On the other hand, extracting implant-specific features remains possible because each convolutional layer has multiple independent filters with parameters adjusted to the data.

\subsection*{Temporal information processing}
We tested five methods to aggregate temporal information. CNN2D+LSTM+MT obtained the highest average cosine similarity. This confirmed that LSTMs could decode ECoG signals into complex hand trajectories. As the hidden state of the LSTM cell was the output of the network, the final predicted trajectory could be influenced by the input data from each time step. 3D CNNs that also analyze temporal information do not have this memory, so convolutional filters in the first layers are not aware of a longer temporal context than the length of the kernel. LSTM's memory can increase the network's ability to predict the desired trajectory as the target position is constant through the analyzed one second of the signal. Therefore, using several even similar target vectors to train the network may create a more robust and precise final estimation. However, the improvement compared to the CNN2D+FC model was not statistically significant, so one can also use this model, which has fewer parameters and performs similarly.

\subsection*{Multiple trajectories}
CNN2D+LSTM+MT model achieved significantly higher cosine similarity than the CNN2D+LSTM. In CNN2D+LSTM models, the LSTM state encodes previous trajectories. It can memorize past hand and target positions and then modify the memory based on the next steps' data. Finally, the state of the LSTM is a summary of past desired trajectories. Providing LSTM with additional information about earlier desired trajectories enables a more accurate representation of the system state with a broader context. LSTMs at each time step decide what to memorize and what to forget. This can be especially useful in the case of repeated imagined MI patterns and varying patient concentration levels, resulting in temporal changes in the level of information contained in the data. Hence, we anticipate that the attention mechanism may be a reasonable way to extend the CNN2D+LSTM+MT model. Attention modules can highlight parts of the input containing the most relevant information for the prediction. Therefore, we expect that models incorporating more advanced attention can further improve hand movement decoding. Another way to extend the LSTM context may be increasing input signal length and taking into account long temporal relationships that occur inside one trial (average trial length $\approx 25$ seconds). Our results show that including even only a one-second-long time series of desired trajectory variation improves decoding accuracy. Having access to previous target variables gives more awareness, to the ML models, about the processes taking place in the experiment. Thus, extending the length of the input signal combined with the attention mechanism could improve overall BCI performance.

\subsection*{Design choices}
We searched for the particular design choices that had the strongest influence on the DL-based models' accuracy. Our results show that the most impactful component of the CNN architecture is the dropout layer. It enabled us to create a significant difference between the accuracy obtained by the MLP and the CNN2D+LSTM+MT. Our dataset has a few samples compared to the number of trainable parameters, so it is especially prone to overfitting. Dropout can be a remedy for overfitting as it is a regularization method \cite{srivastava_dropout}. Surprisingly, the CNN2D+LSTM+MT model achieved a high performance even when the dropout value was 0.65. It corresponds to a strong regularization with more than half of the network switched off. This value is different from the one suggested for computer vision. Cai \etal \cite{cai2020effective} reported a decrease in accuracy for channel-wise dropout rates higher than 0.1. Higher dropout levels limit model capacity and distort information stronger. We hypothesize that the difference in optimal dropout rate between computer vision and ECoG signals originates from the difference in signal to noise ratio but may also be influenced by the network's size. A large network regarding the complexity of the problem can favor stronger overfitting. Therefore, we suggest that it may be possible to reduce the number of convolutional filters and the dropout rate conjointly. This would decrease the number of parameters of the model and reduce training and inference time.

The other design choices had a much smaller influence on the performance. We can notice a slight but general downward trend in performance with these design choices for almost all tested architectures. Our results concluded that ReLU is a better choice than ELU for our problem. This is in contradiction to the conclusions of Schirrmeister \etal \cite{schirrmeister_deep_2017}, who reported a higher decoding accuracy using ELU. However, Schirrmeister \etal \cite{schirrmeister_deep_2017} analyzed raw EEG signals, while our networks were trained from ECoG time-frequency features. Therefore, both analyses consider signals represented in different domains, which certainly influences the choice of architecture.

Regarding pooling method selection, no particular trend was observed when the no-padding option was replaced by max pooling. One of our arguments for the use of no-padding is the specific arrangement of ECoG electrodes. The 64 ECoG electrodes are placed on two $4 \times 4$ cm grids that record neural signals from a small and specific area of the brain. The spatial resolution of the recording is higher than in the case of EEG by orders of magnitude. We see an opportunity for machine learning models to take advantage of this fact. However, max pooling may prevent the detection of small signal variations and make models invariant to small translations. Those may be important due to the nature of the observed phenomenon and the fact that we try to decode precise 3D movements that can be coded in tiny signal variations. Moreover, ECoG implants are centered on the sensorimotor cortex responsible for hand movements. Then, the central features are expected to be the most informative. No-padding naturally directs the attention towards the center of the implants, as fewer convolution operations are performed on the edges of the implants.

In our study, we also analyzed the influence of model depth on cosine similarity. Using more than two convolutional layers can decrease the models' accuracy or give only a slight improvement, depending on the dataset. This result might seem counter-intuitive, as stacking more layers in the DL architectures increases the number of trainable parameters and their capacity for representation. Nevertheless, our input data corresponded to handcrafted features, which might not enable the extraction of high-level features. Worse, increasing the depth of the networks might have resulted in difficulties in estimating the optimal value of the model parameters. This might explain why we do not need to use dozens of layers, like computer vision \cite{he_deep_2015}, for this application. Finally, our best proposed model CNN2D+LSTM+MT consists of four layers---two convolutional blocks and two LSTM layers. Our results are consistent with the vast majority of BCI studies that used DL for EEG \cite{Roy_2019}, ECoG \cite{xie_decoding_2018,du_decoding_2018,elango_sequence_2017, pan_rapid_2018, rashid_electrocorticography_2020} and applied less than 10 layers inside the models.

\section*{Data availability statement}
The data analyzed during the current study are not publicly available for legal/ethical reasons but are available from the corresponding author on reasonable request.

\ack{The authors would like to thank Thomas Costecalde, Serpil Karakas, Felix Martel (all from CEA-Leti), and Stephan Chabardes (CHUGA) for designing and recording the dataset used in this study.

Clinatec is a Laboratory of CEA-Leti at Grenoble and has statutory links with the University Hospital of Grenoble (CHUGA) and with University Grenoble Alpes (UGA). This study was funded by CEA (recurrent funding) and the French Ministry of Health (Grant PHRC-15-15-0124), Institut Carnot, Fonds de Dotation Clinatec. MŚ was supported by the CEA NUMERICS program, which has received funding from European Union's Horizon 2020 research and innovation program under the Marie Sklodowska-Curie grant agreement No 800945. Fondation Philanthropique Edmond J Safra is a major founding institution of the Clinatec Edmond J Safra Biomedical Research Center.}

\appendix

\section{Targets positions} \label{app:targets-positions}
\begin{figure}
	\centering \includegraphics[width=\columnwidth]{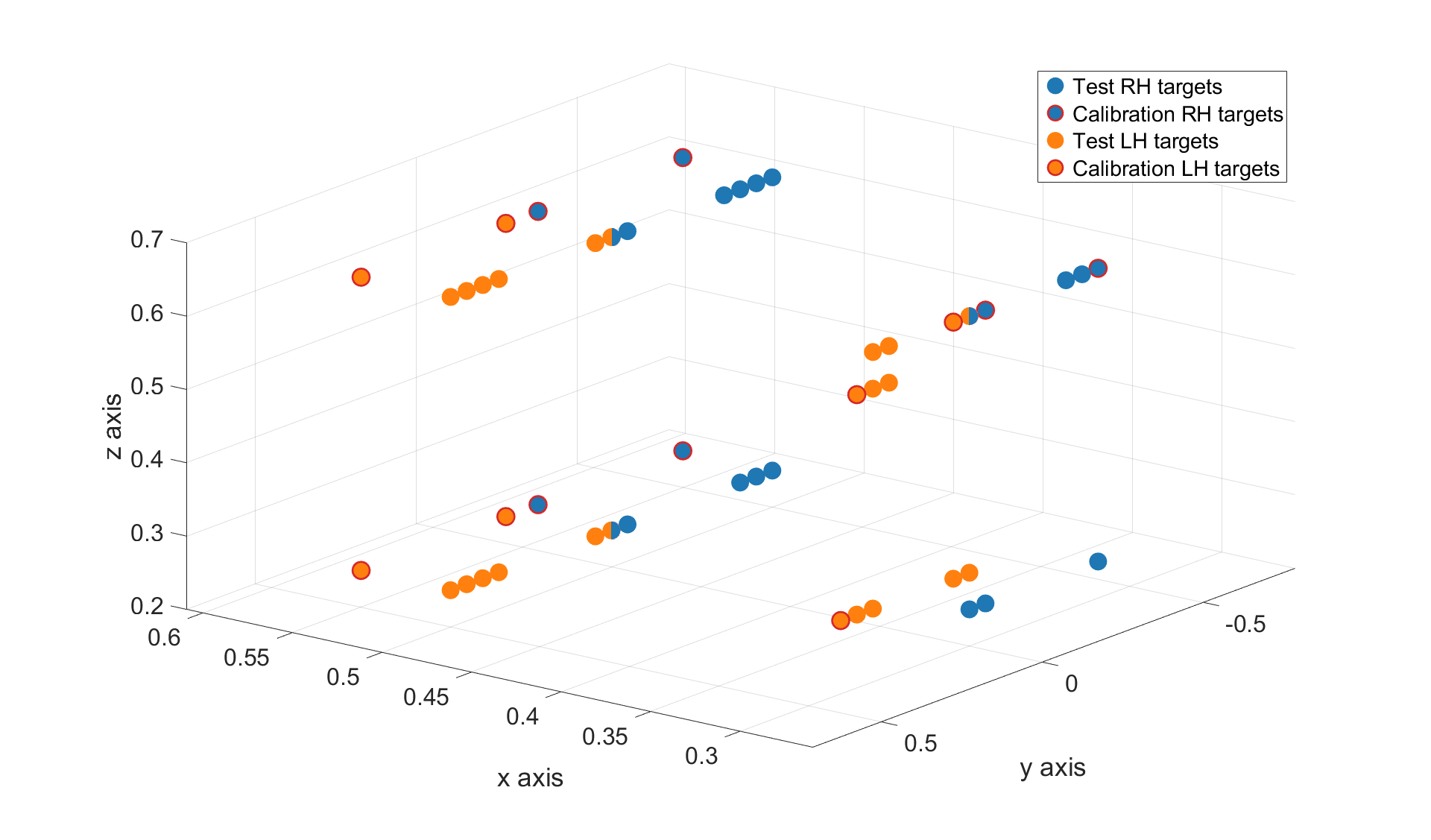}
	\caption{Positions of targets that patient was asked to reach during experiment.} \label{fig:targets-positions}
\end{figure}

In Figure \ref{fig:targets-positions}, we presented positions of targets that were used during calibration and testing sessions.

\section{Number of parameters}
\begin{table}[ht]
	\caption{Number of trainable parameters for DL methods.}\label{tab:num-params}
	\begin{indented}
		\centering
		\lineup
		\item\begin{tabular}{@{}ll@{}}
			\br
			& Parameters \\ \mr
			CNN2D+FC            & \053 635               \\
			CNN3D+FC            & \086 851              \\
			CNN2D+LSTM+MT       & 237 912              \\
			CNN2D+LSTM          & 237 912              \\
			MLP                 & 482 953              \\
			CNN2D+tConv/LSTM+FC & 943 011              \\ 
			\br
		\end{tabular}
	\end{indented}
\end{table}

\section{Hyperparameters values}

\begin{table}[ht]
\caption{Values of hyperparameters selected using TPE.} \label{tab:hyperparams-value}
\resizebox{\columnwidth}{!}{
\begin{tabular}{@{}cccc@{}}
\br
Model               & Learning rate & Weight decay & Batch size \\ \mr
MLP                 & 0.00031       & 0.018        & 592        \\
CNN3D+FC            & 0.00064       & 0.3          & 928        \\
CNN2D+LSTM+MT       & 0.00023       & 0.24         & 96         \\
CNN2D+LSTM          & 0.0089        & 0.12         & 352        \\
CNN2D+FC            & 0.003         & 0.22         & 560        \\
CNN2D+LSTM/tConv+FC & 0.00029       & 0.42         & 96         \\ \br
\end{tabular}
}
\end{table}

\begin{table}[ht]
\centering
\caption{Hyperparameters values used to estimate optimal number of layers.} \label{tab:hyperparams-value-layer}
\begin{tabular}{@{}ccc@{}}
\br
Learning rate & Weight decay & Batch size \\ \mr
0.001         & 0.01         & 200        \\ \br
\end{tabular}
\end{table}

\section*{References}
\bibliography{mybibfile}
\end{document}